\documentclass[5p,twocolumn,times,number]{elsarticle}

\usepackage{amsmath}
\usepackage{widetext}
\usepackage[unicode]{hyperref}
\usepackage{caption}
\usepackage{color}
\usepackage{longtable}

\usepackage{lineno}

\usepackage{graphicx,isolatin1}
\usepackage[squaren]{SIunits}

\usepackage{etoolbox}
\biboptions{sort&compress}

\newcommand{\textroman}[1]{\mbox{\rm #1}}
\newcommand{\gfrac}[2]{\displaystyle\frac{#1}{#2}}

\newcommand{\annee}{\mbox{year}}
\newcommand{\cut}{\footnotesize\textroman{cut}}
\newcommand{\size}{\textroman{p}}

\def\QED{\text{QED}}

 \newcommand{\Red}[1]{\textcolor{red}{#1}}

 \newcommand{\Magenta}[1]{\textcolor{magenta}{#1}}

\journal{Nucl.\ Instrum.\ Meth.\ A }

\begin{document}

\begin{frontmatter}

\title{Polarimetry of $\gamma$-Rays Converting to $e^+e^-$ Pairs with
 Silicon-Pixel-Based Telescopes}

\author[add1]{D.~Bernard\corref{cor}}
\ead{denis.bernard at in2p3.fr}

\address[add1]{LLR, Ecole Polytechnique, CNRS/IN2P3,
91128 Palaiseau, France}

\cortext[cor]{Tel 33 1 6933 5534}

\begin{abstract}
There are serious prospects that the next $\gamma$-ray space mission
could use a telescope based on silicon pixel detectors.
I characterize the potential of such active targets for polarimetry
with gamma-ray conversions to pairs and find it excellent both in
terms of selection efficiency and of effective polarization asymmetry.
\end{abstract}
 

\end{frontmatter} 

\section{Introduction}

Efficient measurements of the high-energy $\gamma$-ray polarization
are eagerly awaited as an additional observable would bring insight on
the processes at work in the $\gamma$ emission of cosmic sources such
as pulsars \cite{Harding:2017tdm,Harding:2017ypk} or blazars
\cite{Zhang:2013bna}.

Even though various experimental schemes to achieve the measurement of
the fraction and of the direction of the linear polarization of a
photon beam have been designed and have been characterized
successfully by the analysis of simulated data and / or of data
collected with a polarimeter prototype on a polarized $\gamma$-ray
beam \cite{deJager:2007nf,Ozaki:2016gvw,Gros:2017wyj}, no such
polarimeter has ever been used on a space mission and therefore no
polarimetry data of cosmic sources in the MeV energy range is
available to date.

Polarimetry is performed by an analysis of the distribution of an
azimutal angle, $\varphi$, that measures the orientation of the final
state in a plane orthogonal to the direction of the incoming photon
(see, e.g., eq. (1) of \cite{Bernard:2022jrj}).
In the presence of $\varphi$ angular resolution, the effective
polarization asymmetry, $A$ is smaller than the theoretical value,
$A_{\QED}$, by a dilution factor $D \equiv A / A_{\QED} $.
One of the main issues is the multiple scattering of the electron and
of the positron in their way through the detector, that makes the
measurement of the azimutal angle of the pair extremely challenging.
In the approximation of a value of the pair opening angle equal to the
most probable value of the distribution of that angle, it was demonstrated 
\cite{Kelner,Kotov,Mattox}
that the dilution factor, $D$, inflicted to the polarization
asymmetry, $A$, varies like
$ D = \exp{(-2 \sigma_0^2 \, x / X_0)} $
with $\sigma_0 \approx 24\, \radian $,
$x$ and $X_0$ the thickness and the radiation length of the
scatterer, respectively.
So for a photon incident normal to and converting at the top of a
$400 \,\micro \meter$ thick silicon wafer, for example,
$D \approx 0.007$, an extremely small value.
It was believed, therefore, that even before the leptons exited the
conversion wafer, most of the azimuthal information was already lost.

This situation led polarimeter designers to extreme decisions, such as
the use of a $100 \,\micro\meter$-thick carbon converter
\cite{deJager:2007nf},
the use of an emulsion active target with a single-grain localization
precision of $60\,\nano\meter$ \cite{Ozaki:2016gvw},
or of a low-density (gas) high-precision active target such as a time
projection chamber (TPC) \cite{Gros:2017wyj}.

A simulation \cite{Bernard:2013jea} using an event generator that samples the
full, 5D, polarized, Bethe-Heitler differential cross section
\cite{May1951} showed, however, that the
decrease of the dilution at high thicknesses is milder than that
predicted by \cite{Kelner,Kotov,Mattox},
see Fig. 17 of \cite{Bernard:2013jea},
as a number of high-opening-angle events
(see Fig. 3 of \cite{Olsen:1963zz}) still contribute to the
sensitivity: there was hope.

An estimate of the performance of a silicon-strip-detector (SSD) /
tungsten-foil active target such as the Large Area Telescope (LAT) on the
{\sl Fermi} mission \cite{Atwood:2009ez}
was obtained from
the analysis of data simulated with a custom-made software and based
on a simplified version of the LAT tracker, with a dedicated event
reconstruction method \cite{Bernard:2022jrj}.
It showed that the effective value of the polarization asymmetry,
$A$, peaks at the small value of 0.02 at a couple of hundred of
MeV (Fig. 21 of \cite{Bernard:2022jrj}).
This value of $A$ is approximately one order of magnitude lower
than the QED value
 (that is, a dilution factor of $D \approx 0.1$).

The variation of $A_{\QED}$ with photon energy can be found in
Fig. 3 of \cite{Semeniouk:2019cwl};
A high-energy asymptotic expression that is valid down to
$\approx 20\,\mega\electronvolt$ was obtained by
\cite{Boldyshev:1972va} (See also eq. (14) of \cite{Bernard:2018hwf}).

In that situation, a sensitive measurement needs a long exposure on a
bright source in the MeV energy range.
A study is in progress for the Vela pulsar, that uses the full
software package of the {\sl Fermi}-LAT Collaboration
\cite{Boinee:2003rk,GLASTLAT:2004vug}
with a simulation using an event generation
\cite{Bernard:2018hwf,Semeniouk:2019cwl,Ivanchenko:2020wvu}
that samples the polarized Bethe-Heitler differential cross-section
\cite{May1951}, and with a dedicated event reconstruction.
The precision, $\sigma_P$, of the measurement of the polarization
fraction, $P$, on the 15 years of data collected by the LAT, was
estimated to be of about 0.15 \cite{Bernard:2022jrj}.
Preliminary results obtained from the analysis of simulated data
\cite{Laviron:2023rhx}
confirm the estimate of the mock-up study \cite{Bernard:2022jrj}.

The LAT has been surveying the $\gamma$-ray sky from 30\,MeV to over
300\,GeV, an energy range covering more than four orders of magnitude, since
2008:
it is just time to plan a successor mission.
Also, between a couple of hundred keV, below which existing Compton
telescopes are quite efficient, and a couple of hundred MeV, above
which the effective area of the LAT is plateauing
\cite{Fermi:LAT:Performance},
lies a ``sensitivity-gap'' for which precise measurements of
$\gamma$-ray sources are sparse.
Several projects are being developed, such as ASTROGAM
\cite{e-ASTROGAM:2017pxr} and AMEGO \cite{AMEGO:2019gny}, that in contrast
to the LAT do not include tungsten converter foils, so the
efficiency and the angular resolution for pair-conversion events at
very low-energies is improved.
Also, in contrast with the LAT for which each detection layer includes
a pair of single-sided silicon-strip detectors, these projects use
layers consisting of a single double-sided silicon-strip detector, so
that both transverse coordinates ($x$ and $y$) of low-energy electrons
can be measured in the same wafer, enabling an efficient detection of
Compton scattering events.

As the sensitivity of the LAT to polarimetry was found to be, to a
large extent, originating from conversions in the silicon part of a
detection layer (i.e., not in the tungsten foil)
\cite{Bernard:2022jrj}, as most of the sensitivity of the LAT was
found to originate from conversions in the ``front'' part of the
tracker, for which the tungsten foils are thinner,
all detectors (LAT, ASTROGAM, AMEGO) can be considered to be ``thin''
detectors.
This implies that 
 the photon conversion probability is much smaller than
unity and the effective area is proportional to the sensitive mass
($28 \, \centi\meter^2 / \, \kilo\gram$ for $100\,\mega\electronvolt$
photon conversions to pairs on silicon with 100\,\% efficiency and
exposure \cite{NIST:gamma}).

As the silicon mass of the three detectors are similar, as the leptons
undergo less multiple scattering in active targets without tungsten
foils, and as ASTROGAM and AMEGO are sensitive to pair conversions at
lower energies, for which the fluxes from cosmic sources
are larger, it can be expected that
the polarimetry performance of the projects be better than that of the
LAT -- even taking into account the different values of the
geometrical parameters of the detector.

Recently AMEGO-X, a modified design of AMEGO, has been proposed with
silicon pixel detectors in place of silicon strip detectors,
with the goal of lowering the capacitance of each segment and therefore to 
enable the detection of lower energy Compton scattering events
\cite{Brewer:2021mbe,Fleischhack:2021mhc,Caputo:2022xpx,Steinhebel:2022ips}.
The limitation of the available electrical power on a space mission
has long hindered the use of pixel-detector active targets for
high-effective area $\gamma$-ray astronomy; 
a way out was found by the use of the low-power Complementary Metal
Oxide Semiconductor (CMOS) technology AstroPix chip recently developed 
\cite{Brewer:2021mbe,Steinhebel:2022ips}.

In this paper I study the polarimetry performance of telescopes using
a pixel-based active target.
I first examine the performance of the proposed AMEGO-X set-up 
\cite{Brewer:2021mbe}, after which I explore part of the parameter space,
the wafer thickness, in particular.
Historically double-sided silicon-strip detectors were designed with
a thickness of $300\,\micro\meter$, \cite{Holl:1987zg,BaBar:1999upx}
while $\gamma$-ray telescopes use thicker wafers
($400\,\micro\meter$ single-sided SSDs for the LAT \cite{Atwood:2007ra},
$500\,\micro\meter$ for ASTROGAM \cite{e-ASTROGAM:2017pxr} and
AMEGO \cite{AMEGO:2019gny})
and while high-energy physicists (HEP) are developing thinner and
thinner pixel detectors, demonstrated at $50\,\micro\meter$ and aiming
down to $20 -40\,\micro\meter$ \cite{ALICEITSproject:2021kcd}.

\section{Method}

This study is performed with similar tools and in a similar way as my
previous work \cite{Bernard:2022jrj}.

The detector consists of a number of infinite planes filling the
``lower'' half space ($z <0$), irradiated from photons that originate
from a cosmic source located ``above'' the detector.
As the selection efficiency and the polarization asymmetry are
observed to decrease strongly at large incident angle with the
detector boresight, $\theta$, events are generated within
$\cos{\theta} > 0.25$.
The orbits of the telescope and the variation of its attitude are
supposed to lead to an isotropic exposure in the detector frame.
The number of planes is large enough that all photons convert.
Pair-conversion event selection and primary Compton-scattering event
rejection are performed from the Monte Carlo information.
The propagation and the interactions of the electrons, of the
positrons and of the photons in the tracker are simulated with EGS5
\cite{Hirayama:2005zm}.

Wafers are of thickness $e = 500\,\micro\meter$ and spaced at a
distance $d = 1\,\centi\meter$, a geometry that corresponds to the
detector described in \cite{Brewer:2021mbe}.
Pixels are defined by a 2D grid of size $\size = 500\,\micro\meter$
\cite{Brewer:2021mbe}, with the ``center'' of the detector,
($x=0$, $y=0$), located at the center of a pixel.
The passive material on the active area of silicon, expected
to amount to $<5\%$ \cite{Brewer:2021mbe}, is neglected.

For each event, pixels having received a strictly positive amount of
energy are recorded.
No discriminating threshold is applied as pixels are used to allow the
detection of keV energy depositions.
Recorded pixels having a side in common are grouped into clusters and
the position of the track at its crossing of that particular wafer is
defined as the geometric barycenter of the
position of the center of the pixels associated to the cluster
(using the barycenter of the energies deposited in each pixel yields
similar results).

\begin{figure}[t]
\begin{center}
 \includegraphics[width=0.935\linewidth]{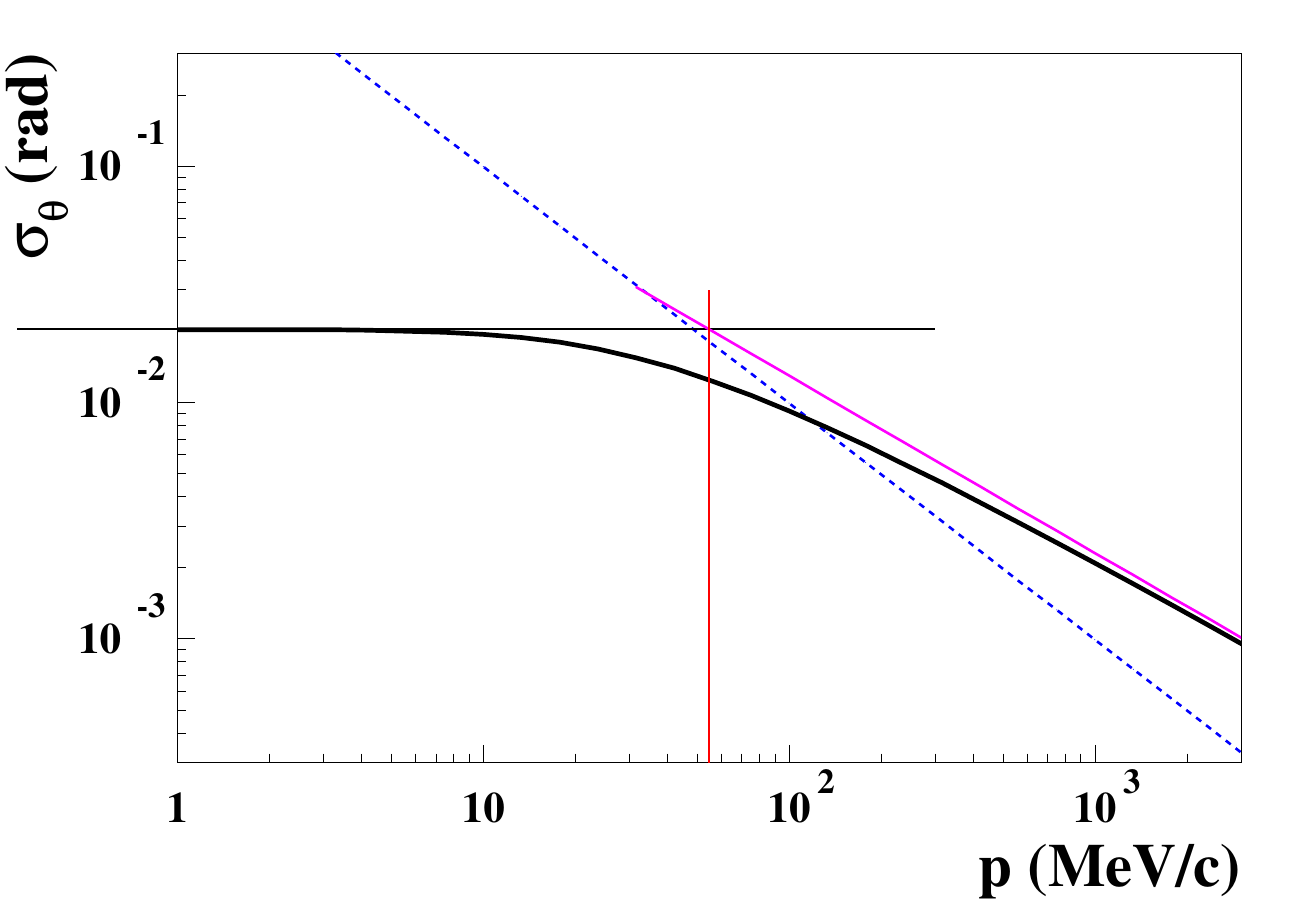}
\put(-110,15){\Red{$p_2$}}
\put(-190,110){$\sqrt{2} \sigma / d$}
\put(-55,80){\Magenta{$\left({p}/{p_1}\right)^{-3/4}$}}
\caption{Single-track polar-angle resolution for a $\gamma$-ray
 conversion at the bottom of a wafer, as a function of track
 momentum.
 The thick curve is the exact solution (eq. (1) of \cite{Bernard:2019znc}).
 The horizontal line is the low-momentum, ``coarse detector'' asymptote.
 The inclined line is the high-momentum, ``homogeneous detector'', asymptote.
 The vertical line shows the limit between both ranges, for $p = p_2$.
 The dashed line shows the RMS multiple scattering angle undergone through a full wafer.
\label{fig:ang}}
\end{center}
\end{figure}

\begin{figure}[t]
\begin{center}
 \includegraphics[width=0.85\linewidth]{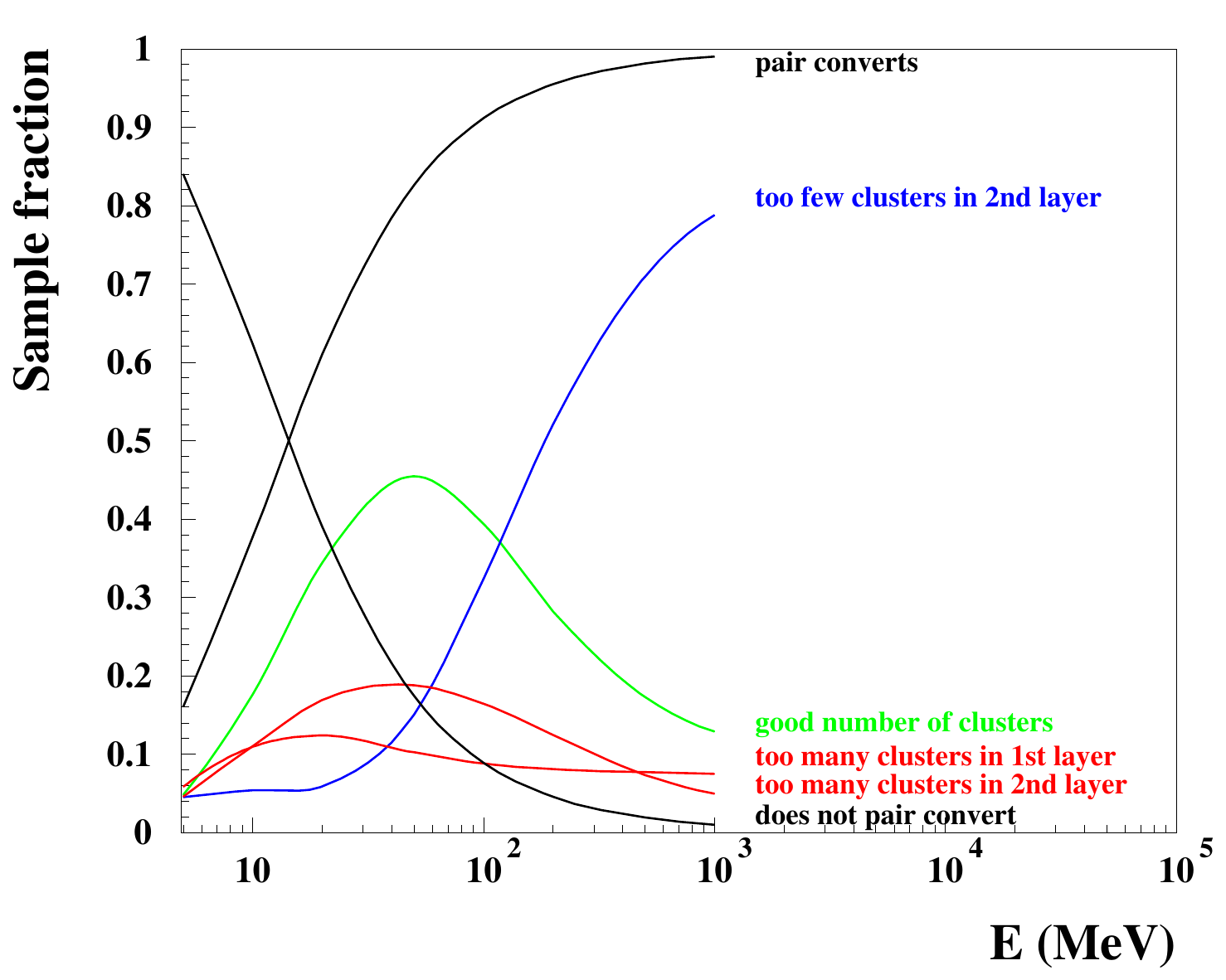}
\caption{Event fractions as a function of incident photon energy.
\label{fig:fraction}}
\end{center}
\end{figure}

\begin{figure*}[t]
\begin{center}
 \hfill 
 \includegraphics[width=0.33\linewidth]{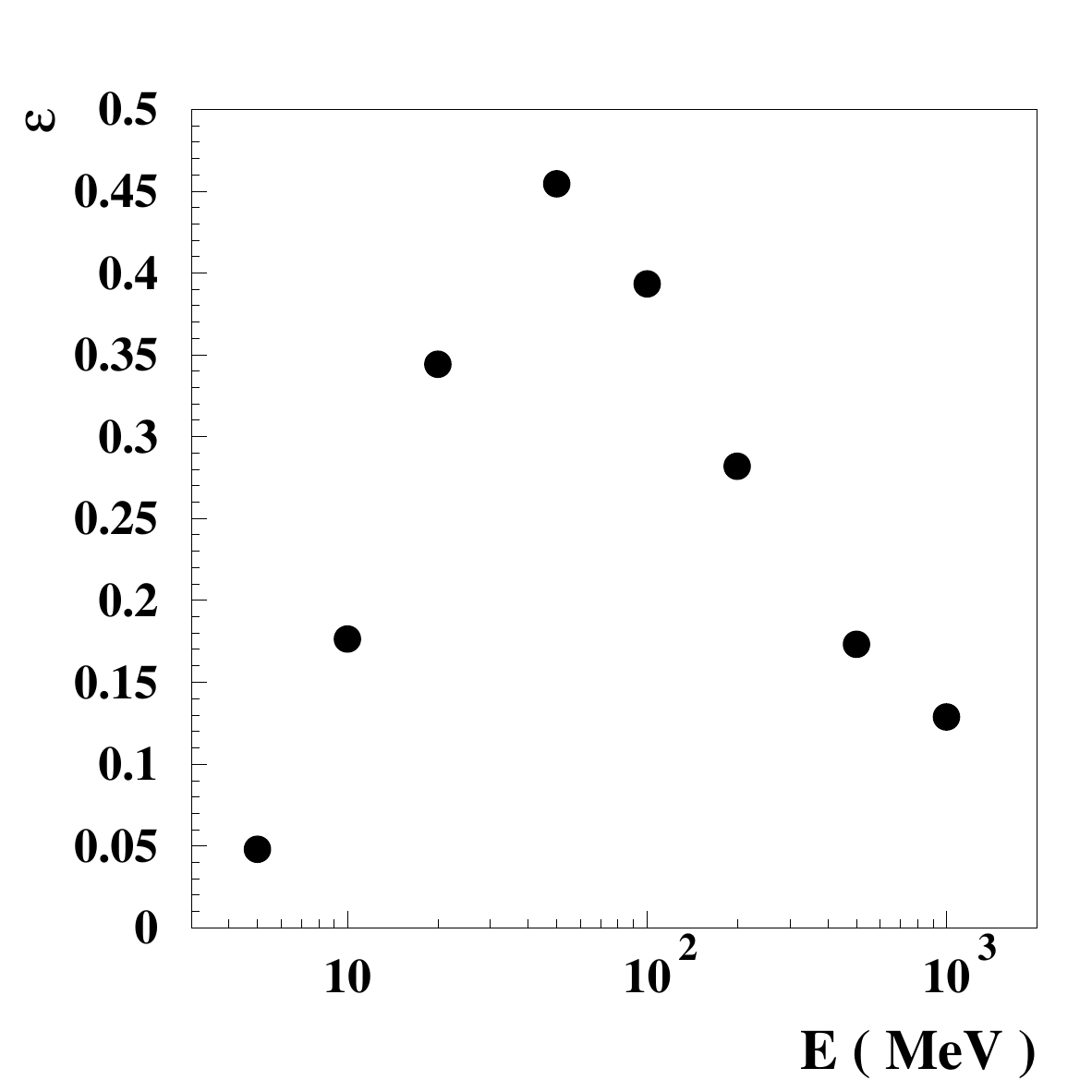}
 \put(-40,130){ {\bf (a)}}
 \hfill 
 \includegraphics[width=0.33\linewidth]{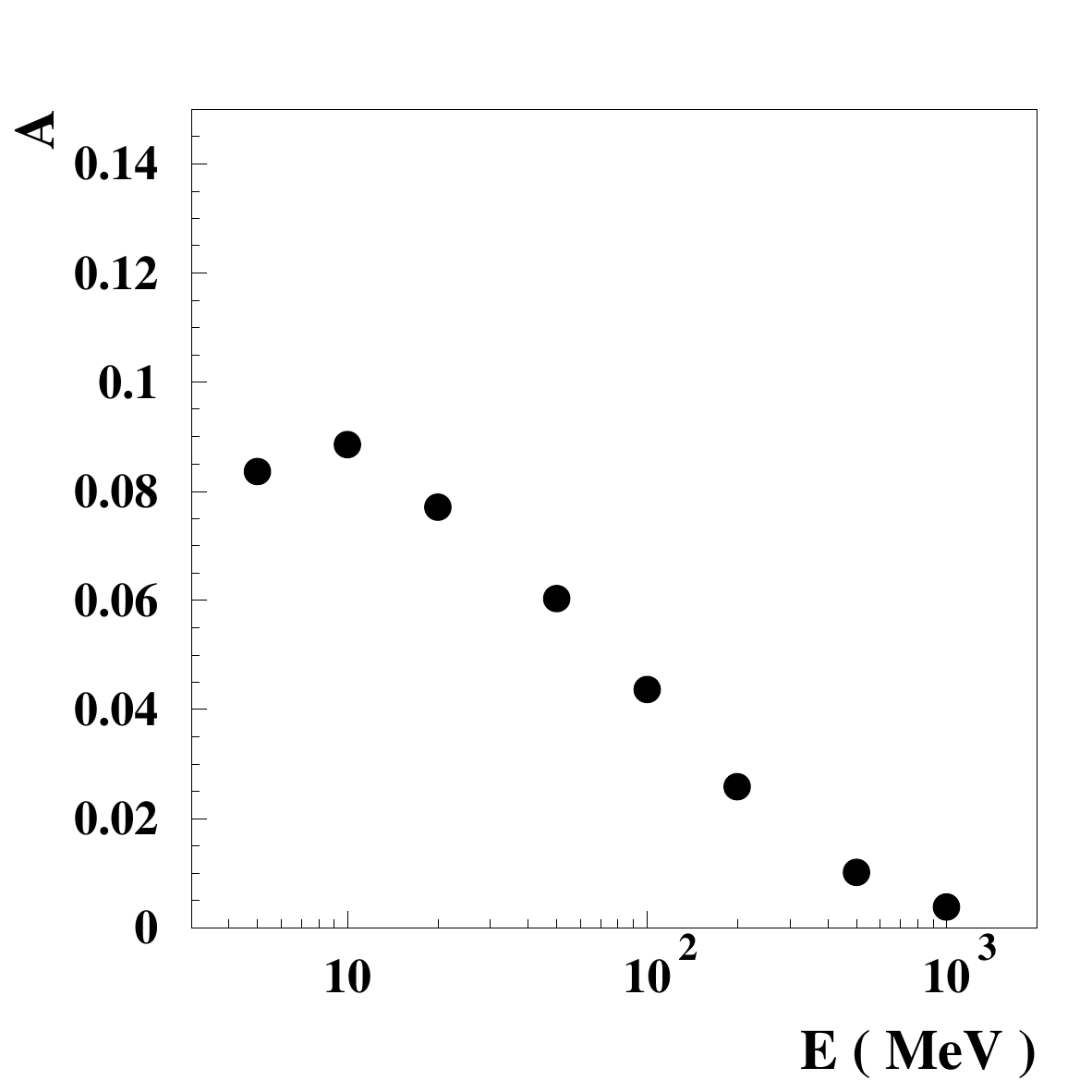}
 \put(-40,130){ {\bf (b)}}
 \hfill 
 \includegraphics[width=0.33\linewidth]{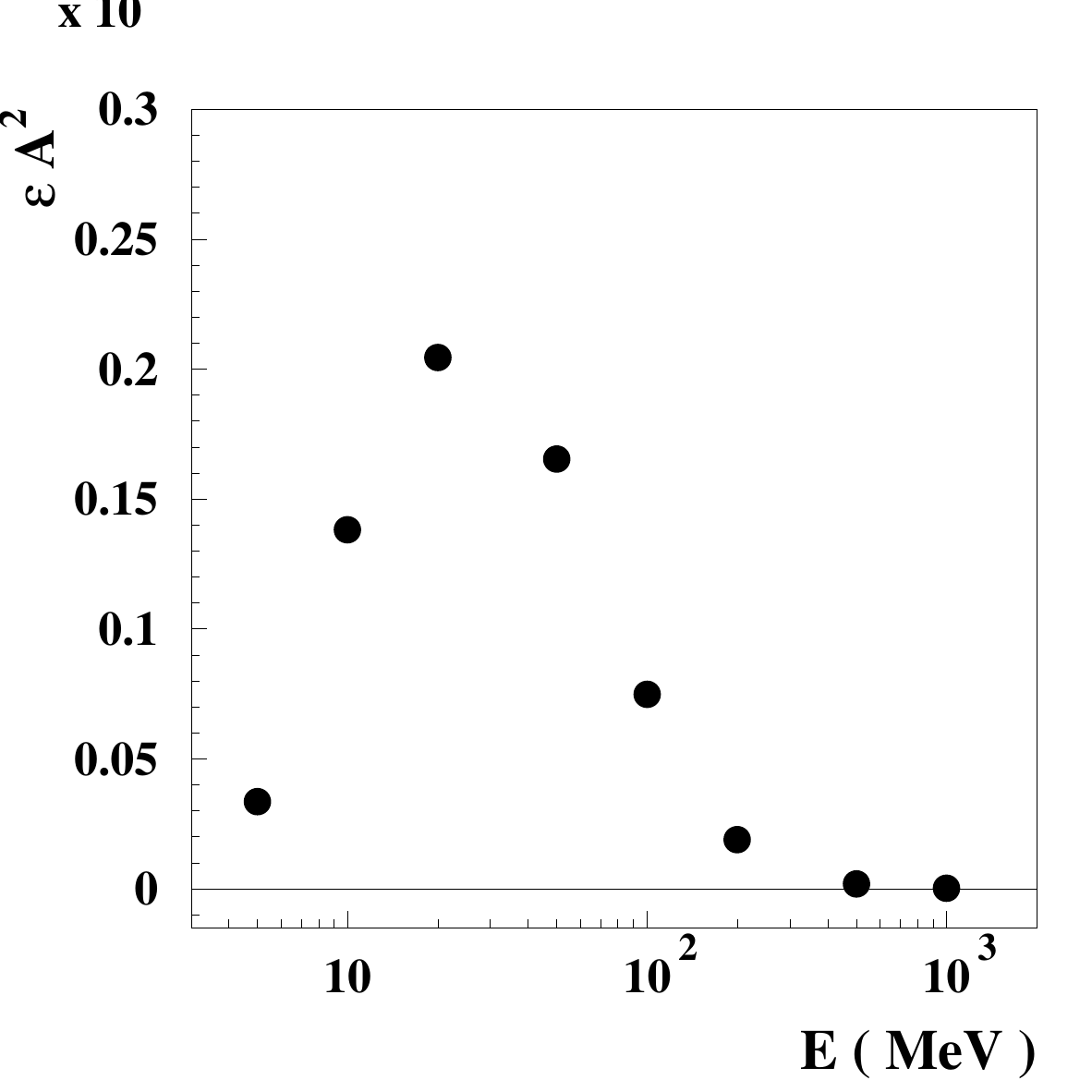}
 \put(-40,130){ {\bf (c)}}
 \hfill 
\includegraphics[width=0.45\linewidth]{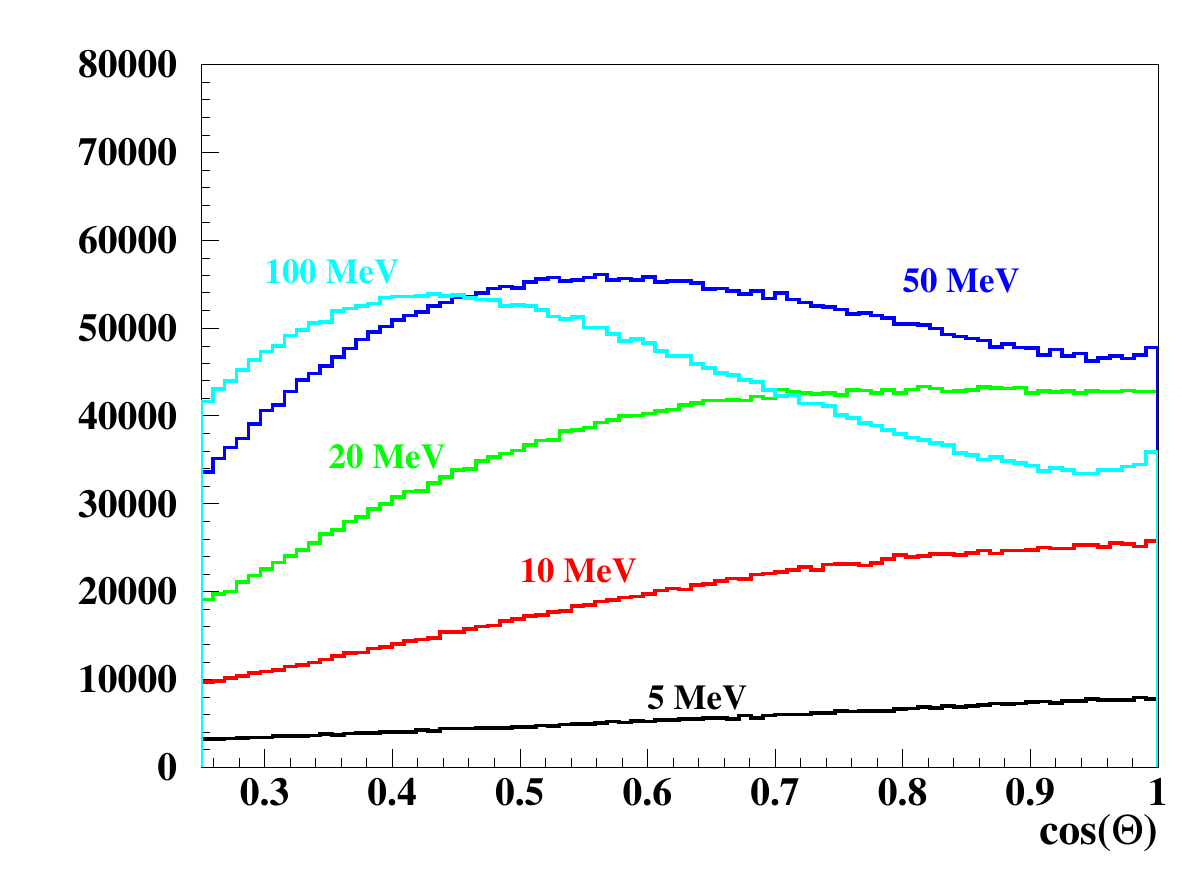}
 \put(-30,140){ {\bf (d)}}
 \hfill 
\includegraphics[width=0.5\linewidth]{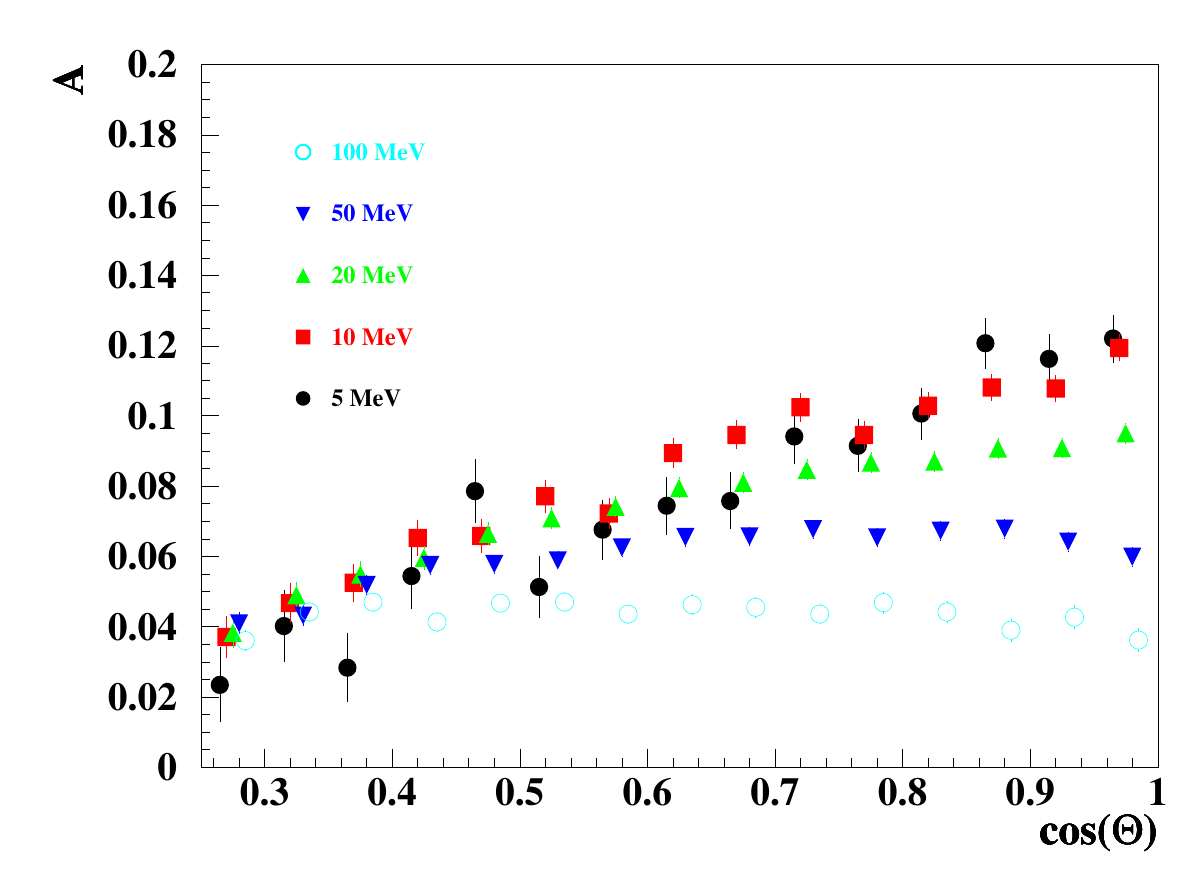}
 \put(-30,150){ {\bf (e)}}
\caption{Simulation of samples of $10^7$ mono-energetic fully
 polarized photons from a cosmic source with isotropic exposure
 within $\cos\theta > 0.25$,
 detected in a simplified version of the pixel
 detector telescope described in \cite{Brewer:2021mbe}.
{\bf (a)} Selection efficiency, $\epsilon$, as a function of incident
 photon energy, $E$.
{\bf (b)} Effective polarization asymmetry, $A$, as a function of incident
 photon energy.
{\bf (c)} Figure of merit, $\epsilon \times A^2$, as a function of incident
 photon energy.
 {\bf (d)} Incident photon $\cos\theta$ distributions for selected
 events, for several photon energies.
{\bf (e)} Effective polarization asymmetry as a function of incident
photon $\cos\theta$ for several photon energies.
Small horizontal shifts have been applied to improve readability.
\label{fig:hop}}
\end{center}
\end{figure*}

Tracking in the presence of detector position resolution and of
multiple scattering is usually performed in an optimal way using a
Kalman filter-based fit \cite{Fruhwirth:1987fm}.
In the case of the LAT, it was shown that over most of the energy
range for which the detector has some sensitivity to polarimetry, the
polar angle angular resolution of such a fit (neglecting the energy
loss of the leptons in their way through the layers), is
asymptotically (at low momentum) equal to the angular resolution
obtained from measuring the angle simply from the track position in
the conversion layer and in the next layer.
Hence a Kalman filter was not used \cite{Bernard:2022jrj}.

The single-track polar-angle resolution obtained from an optimal fit
such as a Kalman-filter fit, for a $\gamma$-ray conversion at the
bottom of a wafer, was studied in \cite{Frosini:2017ftq}.
The R.M.S value is given by eq. (1) of \cite{Bernard:2019znc}, is
drawn as the thick curve of Fig. \ref{fig:ang} and shows two
asymptotes:
\begin{itemize}
\item at low track momenta, the angular resolution is
 $\sigma_\theta\approx \sqrt{2} \sigma / d$, 
that is, the track direction is simply obtained from its positions in
the conversion layer and in the next layer
($\sigma$ is the single-wafer space resolution, here estimated to be
equal to $\size/\sqrt{12}$);

\item at higher momenta, the power of the Kalman filter kicks in,
 additional layers contribute to the measurement, and the angular
 resolution improves
(with respect to $\sigma_\theta\approx \sqrt{2} \sigma / d$), to
 tend asymptotically to the homogeneous-detector value
 \cite{Bernard:2013jea}
$\sigma_{\theta} \approx \left({p}/{p_1}\right)^{-3/4}$, where the
``characteristic'' momentum $p_1$ of the detector is
\begin{equation}
 p_1 = p_0
\left(\gfrac{2 \sigma}{d} \right)^{1/3}
\left(\gfrac{e}{X_0} \right)^{1/2} .
\label{eq:p1}
\end{equation}
\end{itemize}

$p$ is the track momentum and 
$p_0 = 13.6 \, \mega\electronvolt / c$ is the multiple scattering constant.
The limit between the low- and the high-momentum ranges can be defined
by the crossing of the two asymptotes
\begin{equation}
 p_2 = p_0 \sqrt{\gfrac{e}{X_0}} \gfrac{d}{2^{1/3}\sigma},
\label{eq:p2}
\end{equation}
that is, for the detector parameters considered in the present study, 
$p_2 \approx 55 \, \mega\electronvolt /c$.
For a conversion within a layer, the contribution of multiple
scattering within the layer, must be considered in addition
(shown as a dashed line in Fig. \ref{fig:ang}).

As was done for the LAT study \cite{Bernard:2022jrj}, only cluster
configurations with one cluster per primary track and no additional 
background noise are used.
In the case of the present study, this corresponds to one cluster in
the conversion wafer and to two clusters in the next wafer.

The directions of the leptons in the sky frame are calculated from
their directions in the detector frame, after which the azimutal angle
of the event, $\varphi$, is calculated as the average (the so-called
``bissectrix'' \cite{Gros:2016dmp}) of the azimutal angles of the
electron and of the positron.

For event samples simulated with fully polarized photons ($P=1$) and
$\varphi_0=0$, the polarization asymmetry is computed from the moment
of the ``optimal'' weight $w = 2 \cos{2\varphi}$
(\cite{Bernard:2013jea} and references therein)
\footnote{In case the polarization angle of the incoming radiation is
 unknown, a combination of sine and cosine would be used
 \cite{Kislat:2014sdf}.}.

\begin{figure*}[t]
\begin{center}
 \hfill 
 \includegraphics[width=0.33\linewidth]{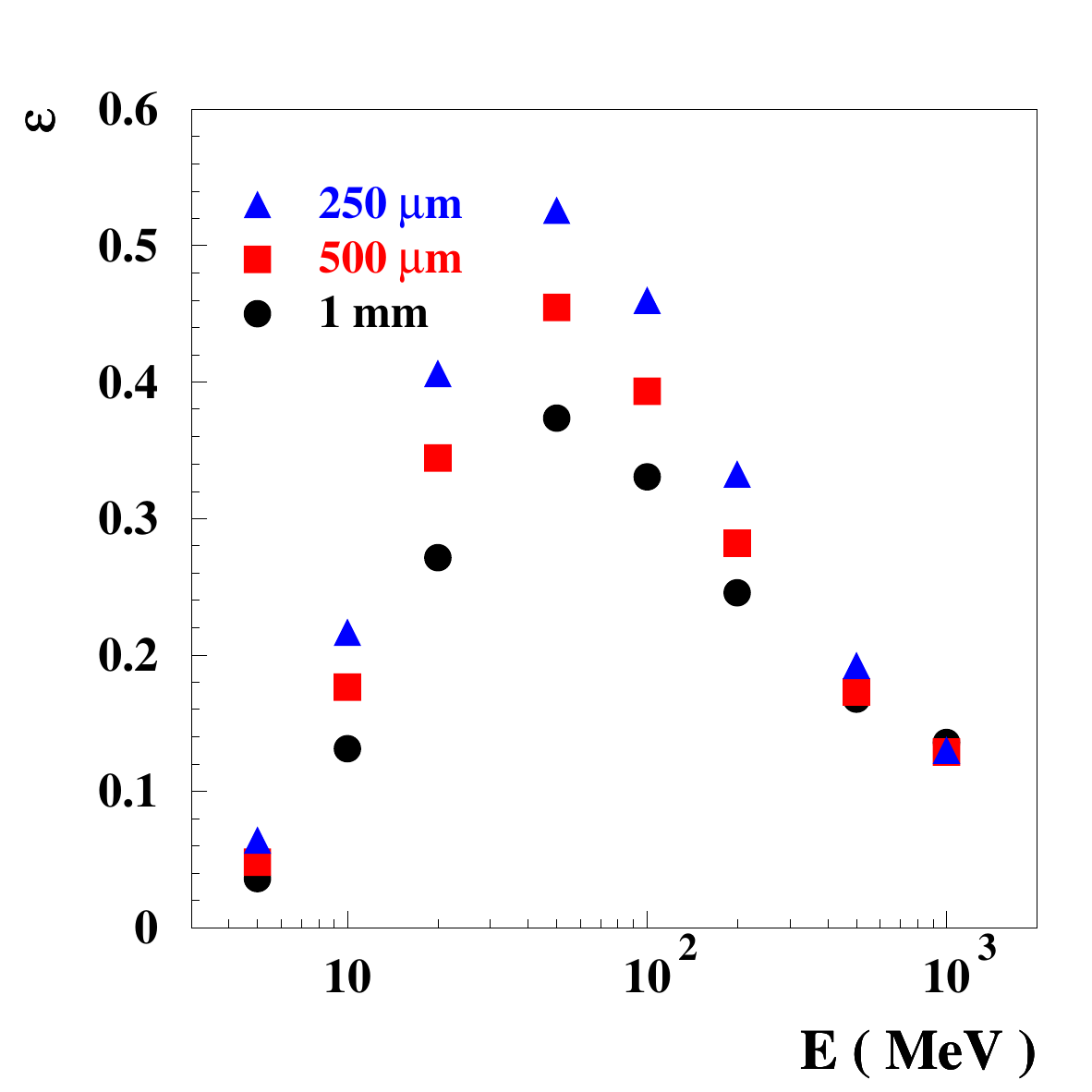}
 \put(-40,130){ {\bf (a)}}
 \hfill 
 \includegraphics[width=0.33\linewidth]{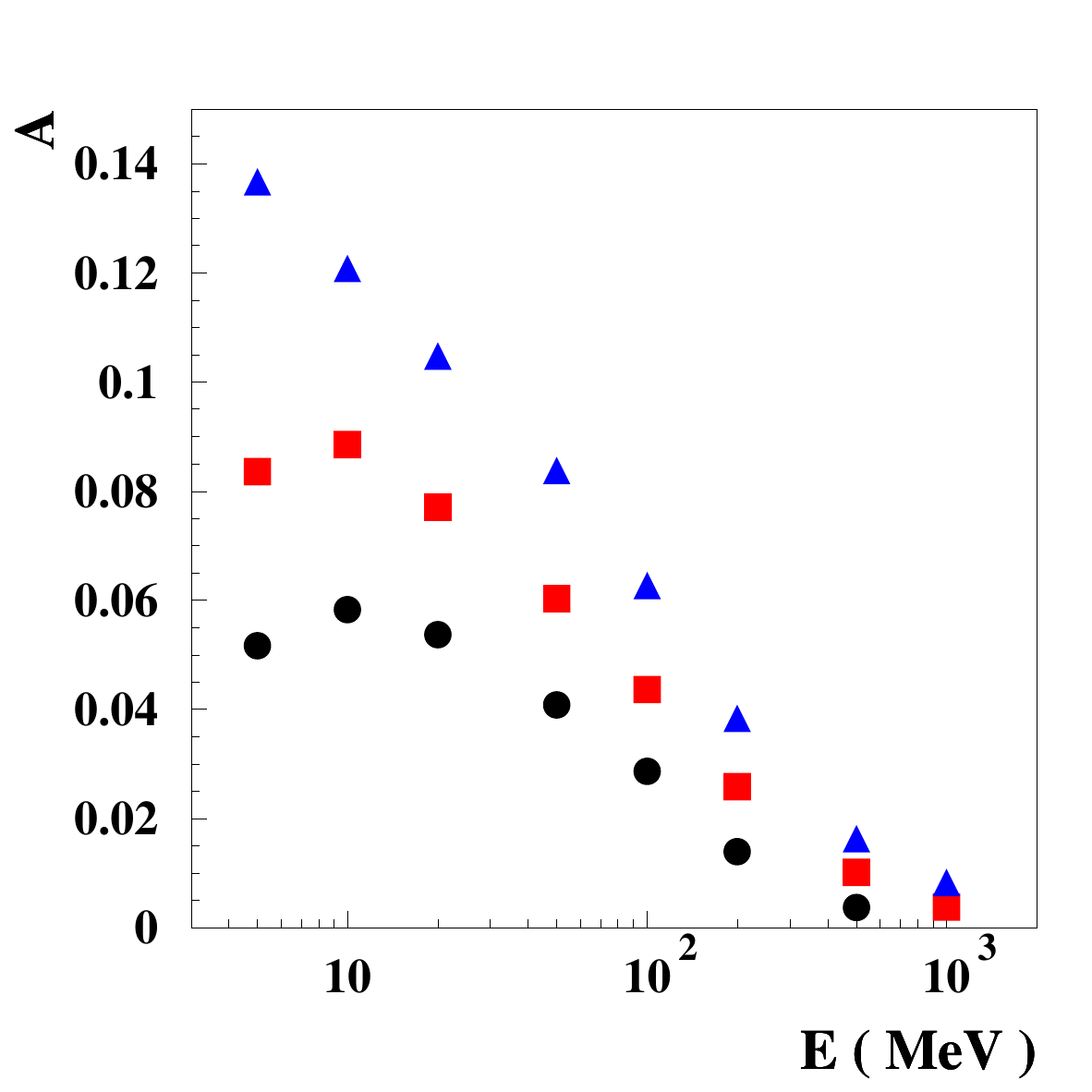}
 \put(-40,130){ {\bf (b)}}
 \hfill 
 \includegraphics[width=0.33\linewidth]{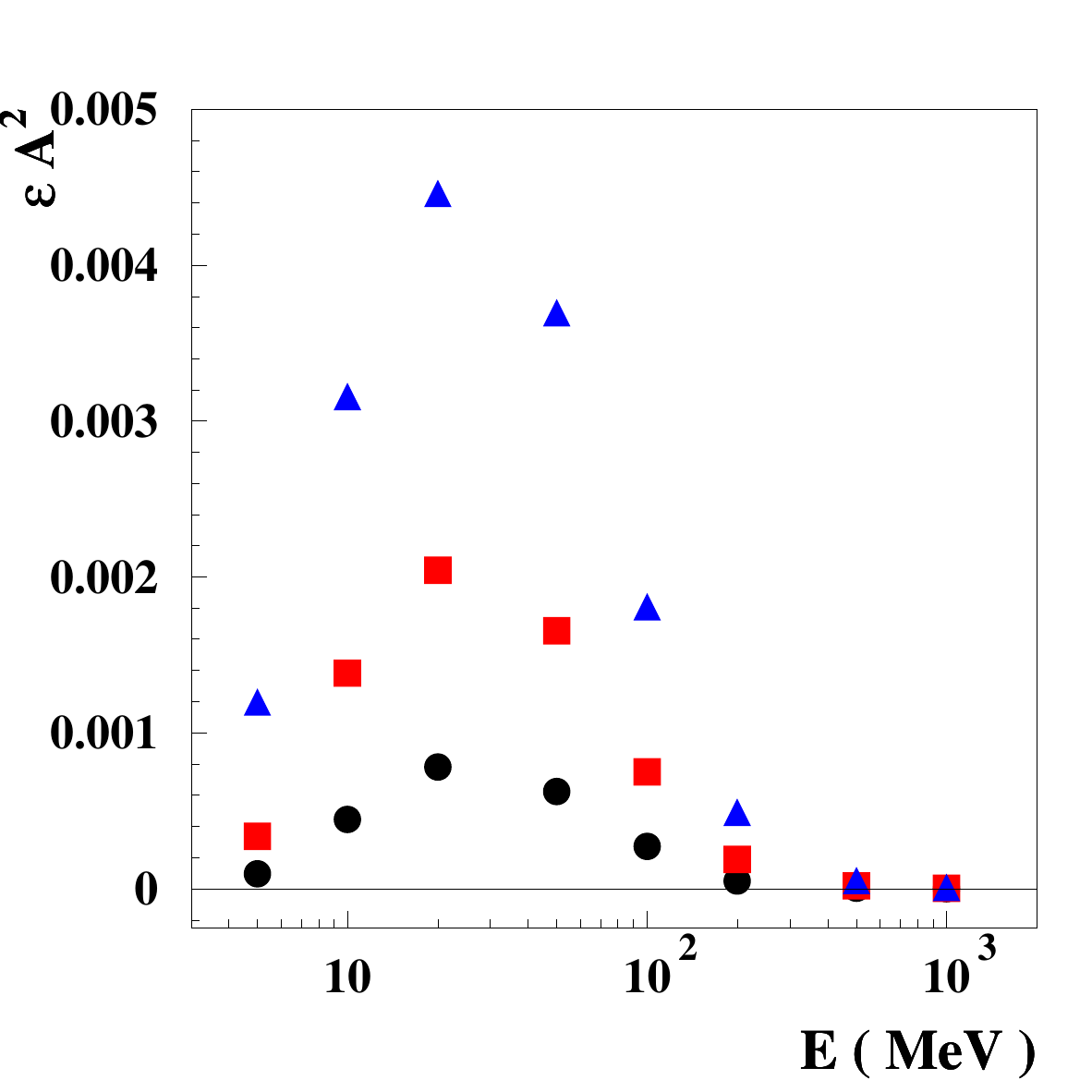}
 \put(-40,130){ {\bf (c)}}
 \hfill 
 \caption{
Study of the variation of the performance of the polarimeter with
wafer thickness, $e$ ($d=1\,\centi\meter, ~ \size=500\,\micro\meter$).
{\bf (a)} Selection efficiency, $\epsilon$, as a function of incident
 photon energy, $E$.
{\bf (b)} Effective polarization asymmetry, $A$, as a function of incident
 photon energy.
{\bf (c)} Figure of merit, $\epsilon \times A^2$, as a function of incident
 photon energy.
\label{fig:t}}
\end{center}
\end{figure*}

\begin{figure*}[t]
\begin{center}
 \hfill 
 \includegraphics[width=0.33\linewidth]{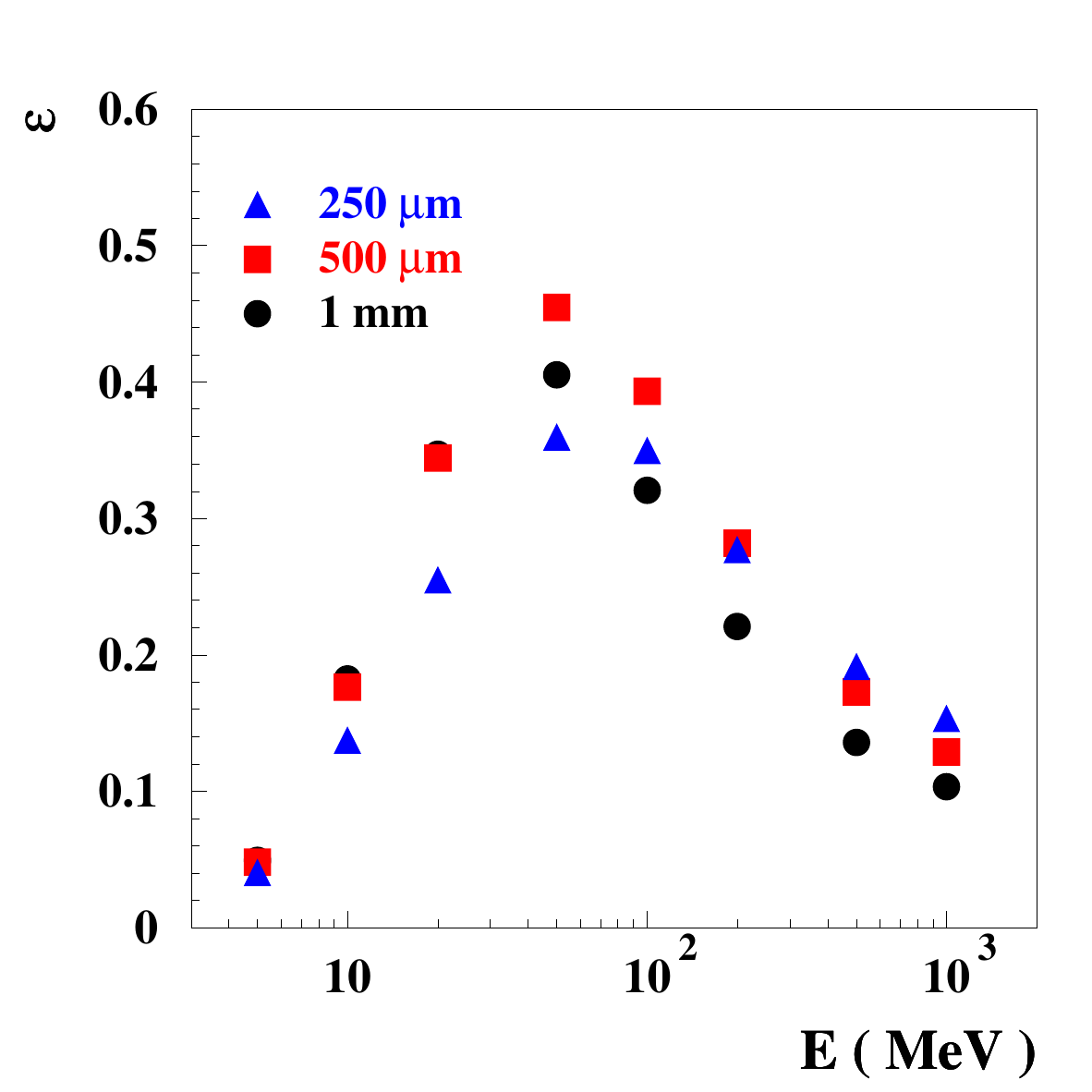}
 \put(-40,130){ {\bf (a)}}
 \hfill 
 \includegraphics[width=0.33\linewidth]{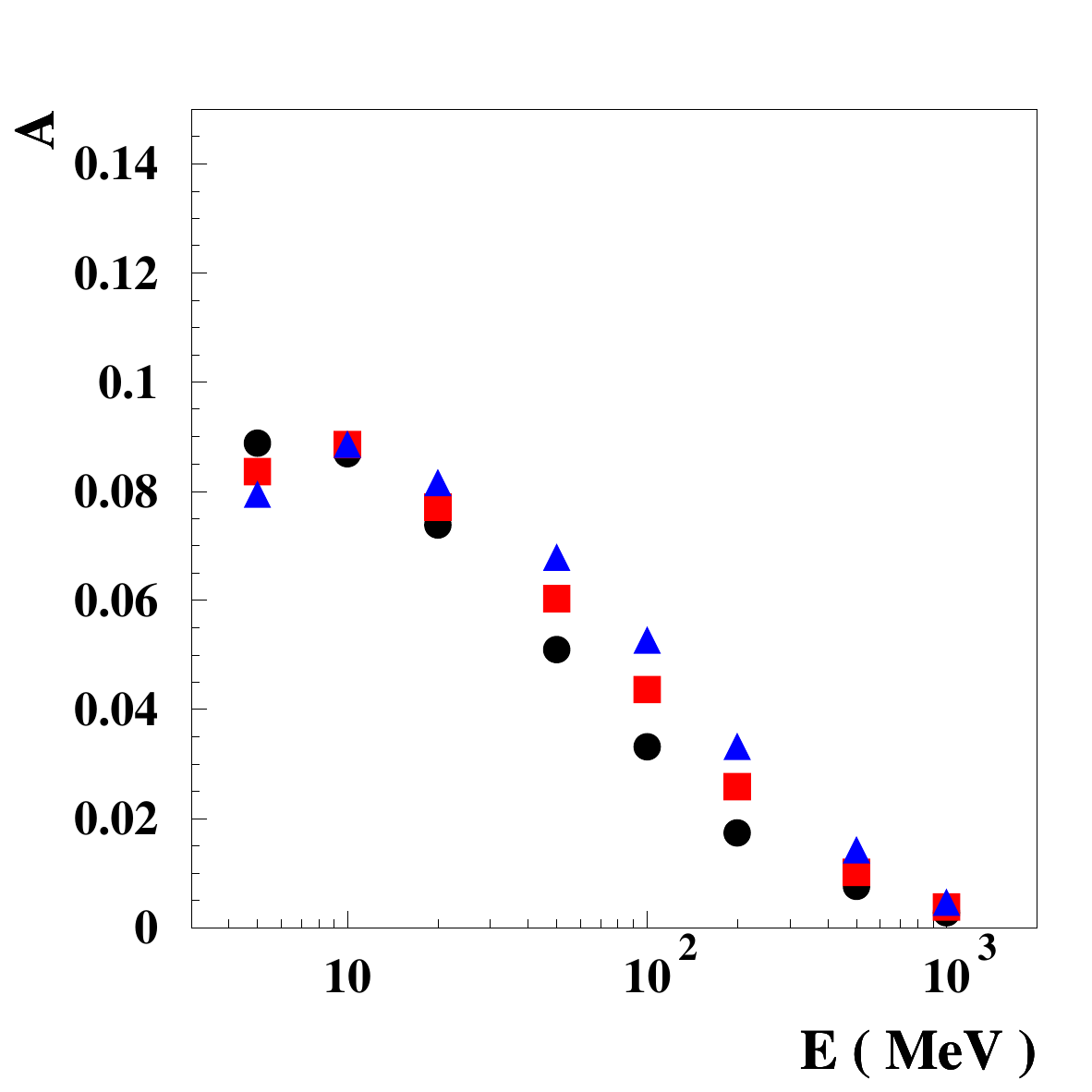}
 \put(-40,130){ {\bf (b)}}
 \hfill 
 \includegraphics[width=0.33\linewidth]{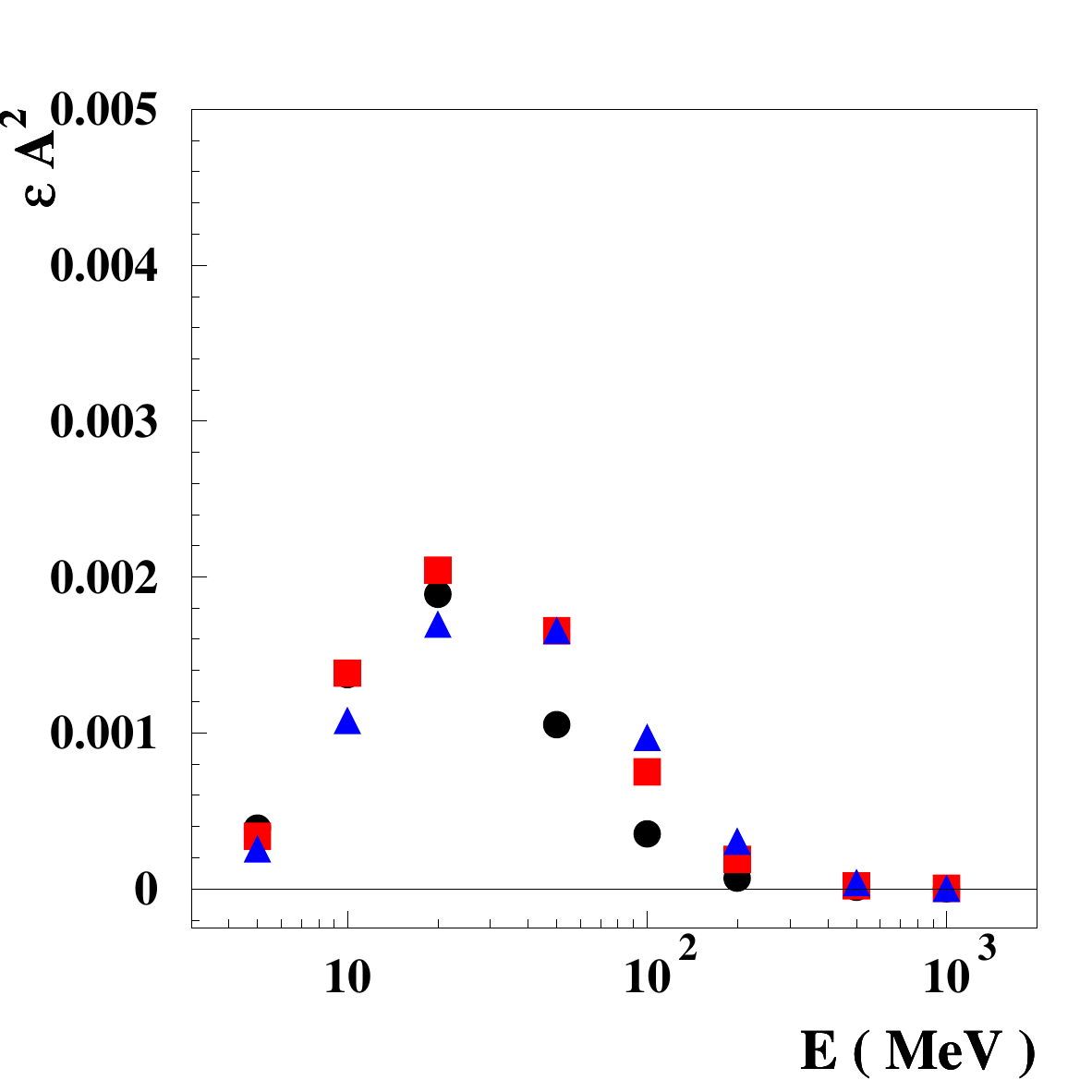}
 \put(-40,130){ {\bf (c)}}
 \hfill 
\caption{
Study of the variation of the performance of the polarimeter with
pixel size, p ($d=1\,\centi\meter, ~ e=500\,\micro\meter$).
{\bf (a)} Selection efficiency, $\epsilon$, as a function of incident
 photon energy, $E$.
{\bf (b)} Effective polarization asymmetry, $A$, as a function of incident
 photon energy.
{\bf (c)} Figure of merit, $\epsilon \times A^2$, as a function of incident
 photon energy.
\label{fig:p}}
\end{center}
\end{figure*}

\begin{figure*}[t]
\begin{center}
 \hfill 
 \includegraphics[width=0.33\linewidth]{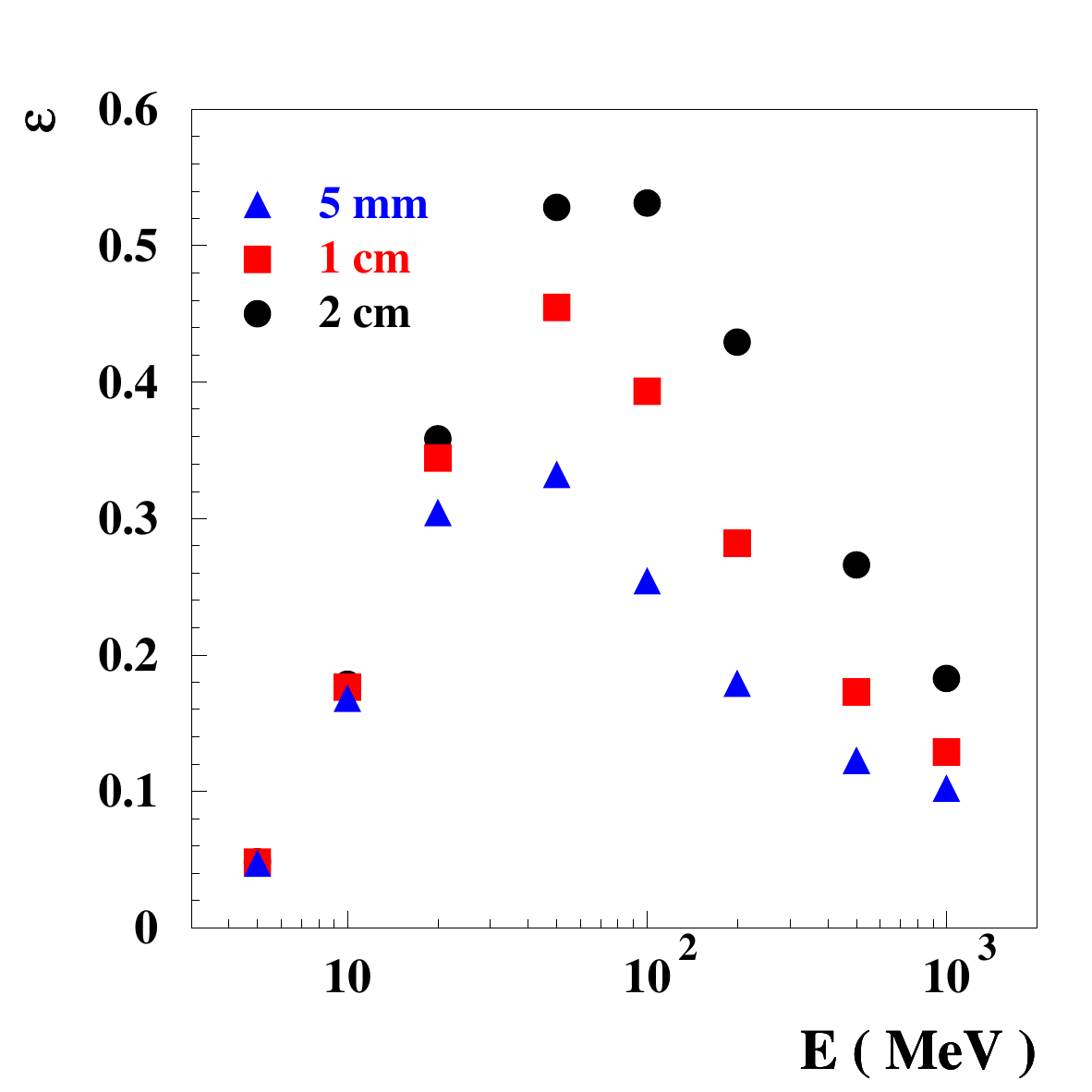}
 \put(-40,130){ {\bf (a)}}
 \hfill 
 \includegraphics[width=0.33\linewidth]{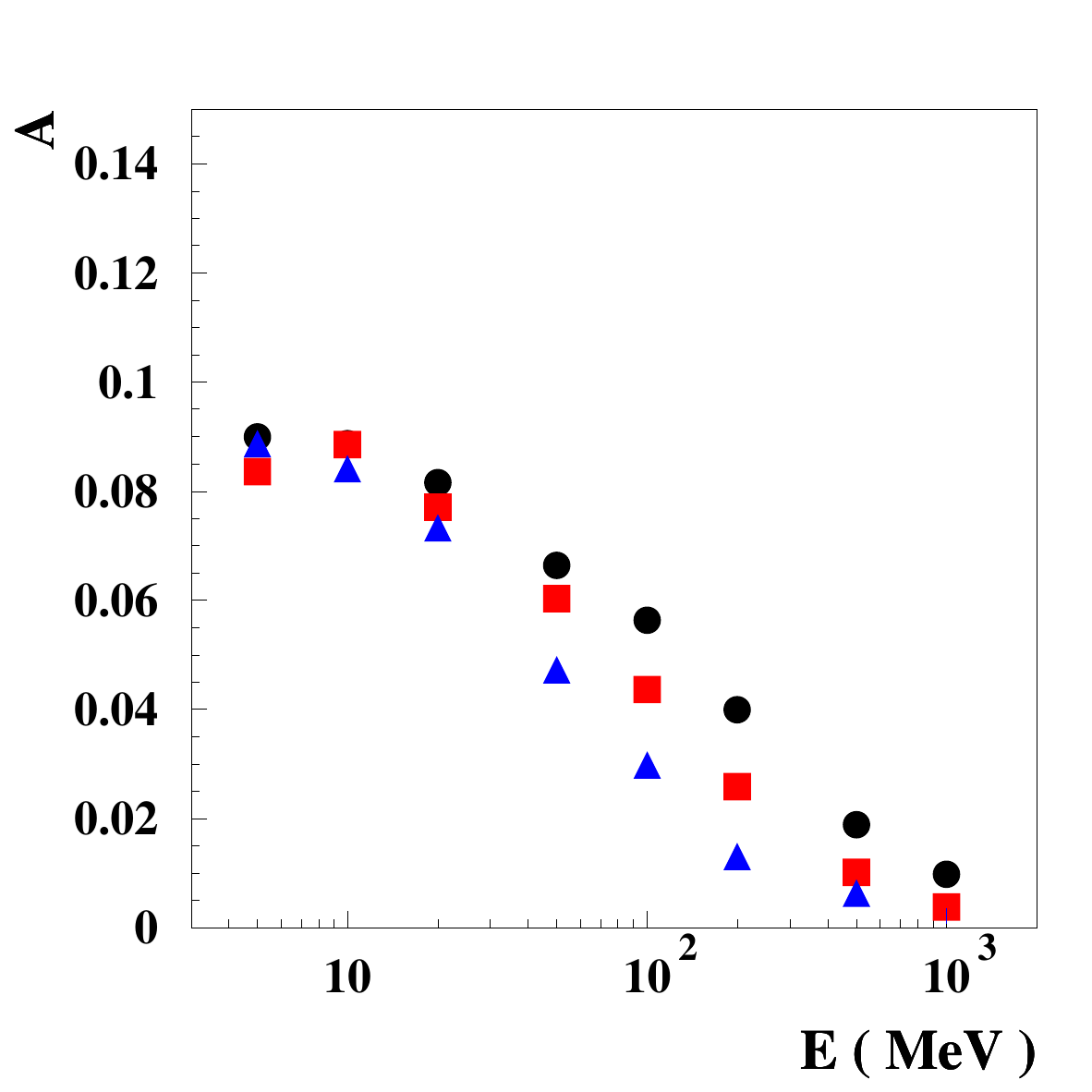}
 \put(-40,130){ {\bf (b)}}
 \hfill 
 \includegraphics[width=0.33\linewidth]{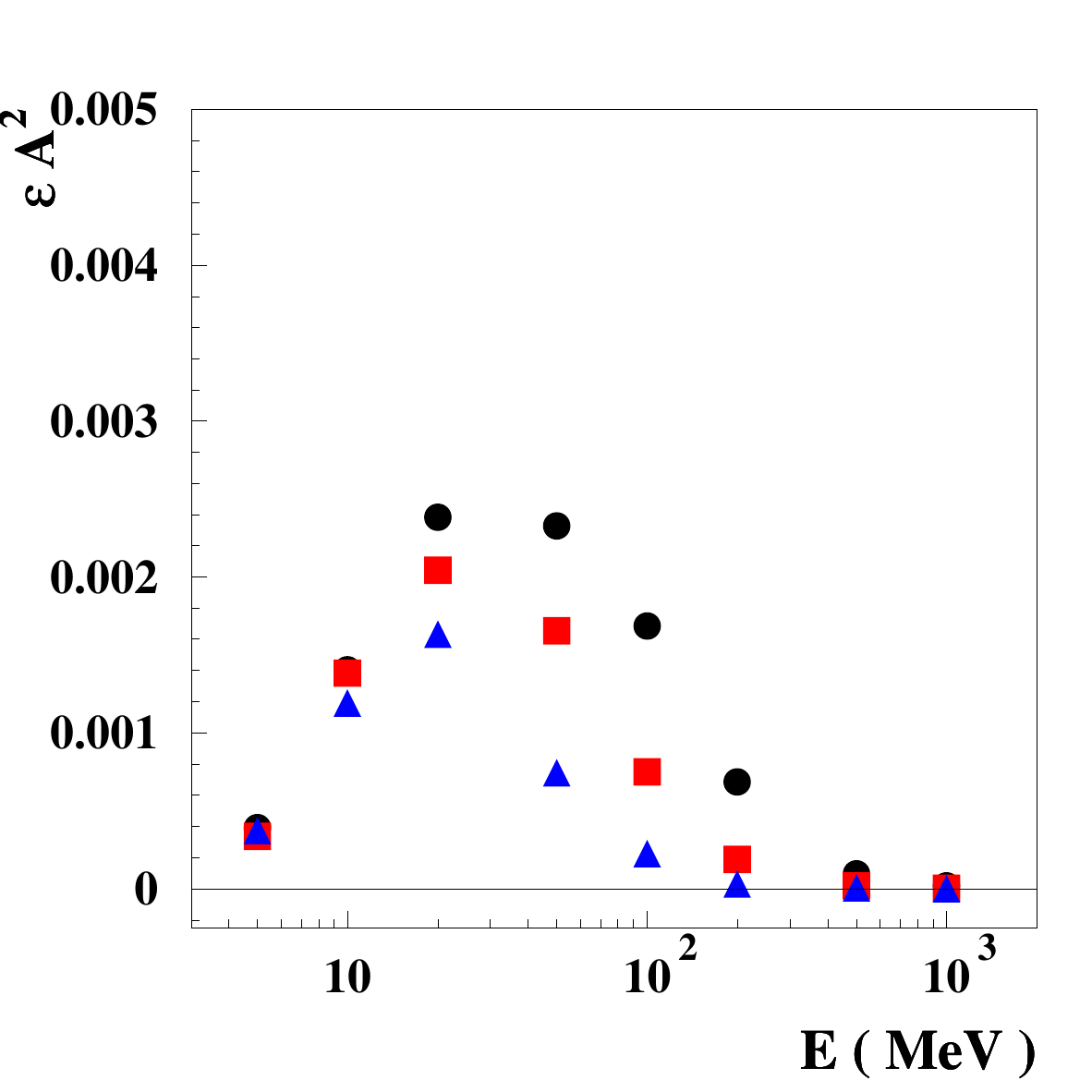}
 \put(-40,130){ {\bf (c)}}
 \hfill 
\caption{
Study of the variation of the performance of the polarimeter with
wafer spacing $d$ $\left(e=500\,\micro\meter, ~ \size=500\,\micro\meter\right)$.
{\bf (a)} Selection efficiency, $\epsilon$, as a function of incident
 photon energy, $E$.
{\bf (b)} Effective polarization asymmetry, $A$, as a function of incident
 photon energy.
{\bf (c)} Figure of merit, $\epsilon \times A^2$, as a function of incident
 photon energy.
 \label{fig:s}}
\end{center}
\end{figure*}

\section{Results}

Samples of $10^7$ mono-energetic fully polarized photons from a cosmic
source with isotropic exposure within $\cos\theta > 0.25$ are
simulated.

The variation of the fraction of events that are relevant to this
analysis (good, too few, too large number of clusters) is presented in Fig. 
\ref{fig:fraction}.
At low energies, a large fraction of events are lost due to Compton
scattering, while at high energies, most events are lost because the
two tracks create only one cluster in the 2nd layer.
Losses due to too many clusters in the 2nd layer are also significant.

The performances of our mock-up polarimeter are presented in
Fig. \ref{fig:hop}.
\begin{itemize}
\item (Plot a), the event selection efficiency, $\epsilon$, is peaking
 at $\epsilon \approx 0.45$ for $E \approx 50\,\mega\electronvolt$
 (the efficiency is larger close to the zenith of the detector,
 small $\theta$, large $\cos\theta$ (plot d)).

\item (Plot b) the effective polarization asymmetry, $A$, is
 plateauing at $A \approx 0.09$
(i.e., a dilution factor of about $1/3$ at 10\,MeV, for which
 $A_{\QED} \approx 0.27$ \cite{Semeniouk:2019cwl})
 at very low energies, and then decreases, to reach $A \approx 0$ at
 $E = 1\,\giga\electronvolt$.

The most probable pair opening angle distribution is peaking at
$1.6\,\radian \cdot \mega\electronvolt /E$, 
\cite{Olsen:1963zz}, that is, at
$1.6\,\milli\radian$ for $E = 1\,\giga\electronvolt$, while the
characteristic sampling angle of the detector, $\size / d$, is equal
to the much larger value of $50\,\milli\radian$
(the so-called critical energy, $E_c$, for which these two angles are
equal \cite{Bernard:2022jrj}, is equal to $32 \, \mega\electronvolt$
for the detector considered here).
$1\,\giga\electronvolt$ photons that converted and that were
detected with such a large opening angle that they managed to create
two separate clusters in the next wafer most likely underwent
 a large multiple scattering deflection, hence the blurring of the
azimutal information
and the small resulting polarization asymmetry.

\item (Plot c), the figure of merit for polarimetry,
 $\epsilon \times A^2$ (see, eg., the discussion in Sects. 12 and 13
 of \cite{Bernard:2022jrj}) is peaking at
 $E \approx 20\,\mega\electronvolt$ and is sizeable mainly between 10
 and $100\,\mega\electronvolt$.
 
\item 
The effective polarization asymmetry, $A$, is obviously larger for
photon originating from the zenith of the detector, (Plot e) and is
plateauing at high $\cos\theta$ for high energies.
\end{itemize}
Most of the sensitivity lies in the energy range
10 -- $100\,\mega\electronvolt$, but keep in mind that this analysis
is using a sub-optimal tracking for track momentum
$p > p_2 \approx 55 \, \mega\electronvolt /c$, that is approximately
for $E > 110 \, \mega\electronvolt $.

I now examine the variation of the performance with the values of the
detector parameters.

~

{\bf Wafer Thickness} ~

\medskip

Increasing the wafer thickness, all other geometrical parameters being
kept unchanged, obviously increases the sensitive (silicon) mass and
therefore the effective area of the telescope.
That main effect is not what I aim to study here, so I generated event
samples simulated with various wafer thicknesses, all samples with the
same number of generated events of $10^7$, which corresponds,
for these ``thin'' active targets, 
to detectors having the same silicon mass.

The residual variation of the performance for polarimetry is presented
in Fig. \ref{fig:t}.
The selection efficiency decreases mildly with increasing thickness (plot a),
mainly due to an increase of the fraction of events
with too large a number of clusters in the 2nd layer, while the
polarization asymmetry obviously decreases with increasing thickness
due to multiple scattering (plot b), and so does the figure of merit
$\epsilon \times A^2$ (plot c).

~

{\bf Pixel Size} ~

\medskip

Decreasing the pixel size obviously improves the selection efficiency
at high energies.
At low energies on the contrary, it degrades it, due to a larger
fraction of events with too many clusters
in the 2nd wafer.
The polarisation asymmetry is barely affected, something that
indicates that the additional contribution of the pixel size to the
$\varphi$ angular resolution is not the dominant factor (Fig. \ref{fig:p}).

~

{\bf Wafer Spacing} ~

\medskip

Decreasing the wafer spacing obviously degrades the selection
efficiency at high energies, as fewer events have a pair opening angle
large enough to induce two separate clusters in the next wafer
(Fig. \ref{fig:s}).
As for the variation of the pixel size, the polarisation asymmetry
is barely affected.

\begin{figure}[t]
\begin{center}
 \includegraphics[width=0.99\linewidth]{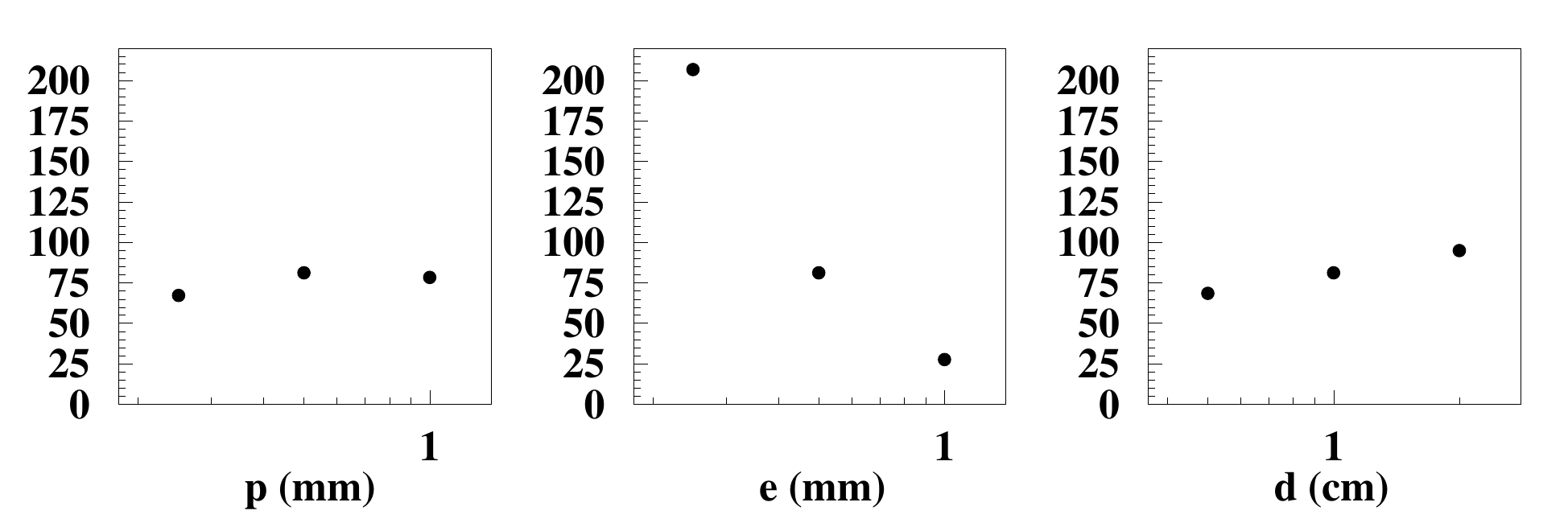}
 \put(-190,60){ {\bf (a)}}
 \put(-110,60){ {\bf (b)}}
 \put(-30,60){ {\bf (c)}}
\caption{Variation of the performance of the polarimeter, expressed as the value of $\lambda \equiv \sum_k A_k^2 N_k / (T \, m)$, with the values of the detector pixel size {\bf (a)}, wafer thickness {\bf (b)}, and wafer spacing {\bf (c)}.
\label{fig:variation}}
\end{center}
\end{figure}

\section{Polarimeter performance}

Given the strong variation of the polarization asymmetry with energy,
the polarization fraction is measured with an event weighting as
described in Sect. 12 of \cite{Bernard:2022jrj}.
The precision of its measurement, $\sigma_P$, is then (eq. (19) of
\cite{Bernard:2022jrj})

\begin{equation}
 {\sigma_P} = 
\sqrt{ \gfrac{2}{\sum_k A_k^2 N_k }}
 .
 \label{eq:uncertainty:comb}
\end{equation}
where $A_k$ and $N_k$ are the effective polarization asymmetry and the
number of events of energy bin $k$, with

\begin{equation}
 N_k =
\gfrac{\eta \, \epsilon_k \, F_0 \, \Delta E_k \, H(E_k) \, T \, m }{E_k^2} , 
 \label{eq:Nk}
\end{equation}

for a bright source with spectral index equal to 2, and a differential flux
equal to $F_0 / E^2$,
$F_0 = 10^{-3}\,\mega \electronvolt /(\centi\meter^2 \second)$,
$m$ the detector silicon mass, 
$T$ the duration of the data collection,
$H(E_k)$ the total photon attenuation at photon energy $E_k$ \cite{NIST:gamma},
$\Delta E_k$ the width of energy bin $k$
and the exposure factor
$\eta = (1 - \cos(\theta_{\cut}))/2 \approx 0.375$.
An overall figure of merit for the performance of a given detector
configuration can be defined as
$\lambda \equiv \sum_k A_k^2 N_k / (T \, m)$.

For the parameter set of this study
($e=500 \, \micro\meter$, ~ $d=1\,\centi\meter, ~ \size=500\,\micro\meter$),
$\lambda$ is found to be equal to
$81. (\annee \cdot \kilo\gram)^{-1}$, so 
for a 5\,year, 30\,kg \cite{Caputo:2022xpx} mission, a precision of
$\sigma_P \approx 0.013$ would be within reach.

The variation of $\lambda$ with the detector parameter values
considered in the previous section is presented in Fig.
\ref{fig:variation}.
The wafer thickness is clearly the parameter that determines the
overall performance of the polarimeter.

\section{Conclusion}

I have studied the performance of an all-silicon, pixel-based, active
target tracker for polarimetry in the pair creation regime.
The analysis proved to be much simpler with pixel detectors than that
for strip detectors \cite{Bernard:2022jrj}.
By comparison with the {\sl Fermi} LAT, the absence of tungsten foils
extends the sensitivity range to lower energies, for which the flux of
cosmic sources is higher and the pair opening angle is larger.
As a result the precision of the polarization fraction that can be
expected is improved, even for a smaller silicon sensitive mass and a
shorter mission duration.
Using pixels (AMEGO-X) instead of strips (AMEGO), with other parameter
(wafer thickness, strip pitch / pixel size, layer spacing) kept
unchanged, removes the loss in polarization asymmetry induced for the
strip detector by the ambiguity between two photon candidates when two
clusters are present in both $x$ and $y$ direction.
The selection efficiencies and the polarization asymmetries are
found to be sizeable.
The selection efficiency improves, at high energies, with larger wafer spacing.
The polarization asymmetry improves with thinner wafers on the whole
energy range.
This study indicates that measurements of the polarization fraction of
the brightest sources of the MeV $\gamma$-ray sky with a precision of
several per-cents may be within reach.

\section{Acknowledgements}

I would like to pay tribute to my colleague Berrie Giebels who
converted me to polarimetry and who sadly passed away last year.


\end{document}